\documentclass[aps,prl,twocolumn,showpacs,superscriptaddress ]{revtex4}

\usepackage{epsf}
\usepackage{bm}

\begin{document}

\title{Nonpolar optical scattering of Positronium in Magnesium Fluoride}

\author{I.V.~Bondarev}\email[Corresponding author.
E-mail: ]{bondarev@tut.by}\affiliation{Institute for Nuclear Problems,
Belarusian State University, Bobruiskaya Str.11, 220050 Minsk,
BELARUS}\affiliation{Laboratoire d'Annecy-le-vieux de Physique des
Particules, BP 110 74941 Annecy-le-Vieux C\'{e}dex, FRANCE}
\author{Y.~Nagai}\affiliation{Institute for Materials Research, Tohoku University,
Oarai, Ibaraki, 3111-1313, JAPAN}
\author{M.~Kakimoto}\altaffiliation[Present address:~]{Communication \& Information System Labs,
R \& D Center, Toshiba, 1 Komukai, Toshiba-Cho, Kawasaki, Kanagawa
212-8582, JAPAN}
\author{T.~Hyodo}\affiliation{Institute of Physics, Graduate School of Arts and
Sciences, University of Tokyo, 3-8-1 Komaba, Meguro-ku, Tokyo
153-8902, JAPAN}

\begin{abstract}
We report the results of the analysis of the temperature
broadening of the momentum distribution of delocalized Positronium
(Ps) in Magnesium Fluoride in terms of optical
deformation-potential scattering model (long-wavelength optical
phonons).~The Ps optical deformation-potential coupling constant
$D_{o}$ in MgF$_{2}$ has been determined to be
$(1.8\pm0.3)\times10^{9}$~eV/cm.~We also show that the Ps momentum
distribution is sensitive to second-order phase transitions in
those crystals where optical deformation-potential scattering is
allowed in one and forbidden in another crystalline phase.
\end{abstract}

\maketitle

Magnesium Fluoride, which is commonly used for optical elements of
extreme ruggedness and durability in both the infrared and
ultraviolet (see, e.g.,~\cite{Blanc}), has recently been proven to
be a good host material for three-dimensional atom optical
lithography on the nanometer
scale~\cite{Konstantz1,Konstantz2}.~In this technique, the
resonance atom-light interaction of dopant atoms is utilized to
structure their positions inside the host MgF$_{2}$ matrix and
thus to engineer electronic and photonic features (e.g., a
photonic band) of a composite solid.~In such nanostructural
applications, it is important to understand vibrational properties
of a host material as they are those disturbing electronic and
photonic bands of nano-composite via deformation-potentials.~The
latter are caused by acoustic and optical strains created by
acoustic and optical vibrational modes, respectively.~These
strains influence the particle in its band in two distinct ways
(see, e.g., Refs.~\cite{Ridley,Bir}).~In the first way,
short-range disturbances of the periodic potential cause
practically instantaneous changes in energy, and these are the
ones quantified by deformation-potentials and referred to as
deformation-potential scattering (acoustic and nonpolar optical,
respectively). In the second way, the distortion of the lattice
may destroy local electric neutrality, and produce electric
polarization and associated macroscopic comparatively long-range
electric fields to which the particle responds.~Disturbance of the
particle's motion by this effect is referred to as piezoelectric
scattering, if associated with acoustic modes, and polar optical
scattering, if associated with optical modes.~Because of its
electric neutrality, the Positronium atom
[$\,\mbox{Ps}\!=\!(e^{+}e^{-})\,$] is less sensitive to the latter
two~\cite{Boev}, thus providing a unique opportunity to study
short-range deformation-potential interactions only.

Physical nature of Ps states in crystalline dielectrics has been
of a continued interest for more than three
decades~\cite{Brandt,Dup,Aref,Boev,Hy88,Bo98,Bo98a,Na00,Gessmann01,Bo01,Hyodo031,Hyodo032,Bo04}
ever since Brandt~\emph{et al.} first identified the delocalized
Ps state in the Angular Correlation of Annihilation Radiation
(ACAR) spectrum of the single crystal of
$\alpha$-SiO$_{2}$~\cite{Brandt}.~Positronium has been found to
form in a delocalized Bloch-type state in dielectric crystals with
low enough concentration of defects (no more than $10^{15}$
defects per cm$^{3}$~\cite{Aref}) at sufficiently low temperatures
(typically below a few tens~K)~\cite{Aref,Hy88}.~The formation of
Bloch-type Ps is confirmed by observing very narrow peaks (the
central peak and satellite peaks appearing at the momentum
corresponding to the reciprocal lattice vectors of the sample
crystal) in the ACAR spectra resulting from the $2\gamma$-decay of
Ps and representing its momentum distribution in a~crystal.~As
temperature increases, it is observed that the central Ps peak
becomes drastically wider and the satellite peaks disappear,
indicating the localization of Ps~\cite{Hy88}.~Such an effect of a
thermally activated self-localization (self-trapping) of Ps was
observed in a~number of dielectric crystals and was studied
theoretically in Refs.~\cite{Bo98,Bo98a} in terms of the
interaction of Ps with the field of a short-range acoustic
deformation-potential (long-wavelength longitudinal acoustic
phonons).

The Ps atom in the MgF$_{2}$ and $\alpha$-SiO$_{2}$ crystals is
known to form a delocalized state in a wide temperature range from
$\sim\!10$~K up to $\sim\!700$~K~\cite{Na00}.~In MgF$_{2}$, the
central and satellite Positronium ACAR peaks are observed to be
drastically broadened above $200$~K.~This extraordinary broadening
was failed to be explained in terms of Ps acoustic
deformation-potential scattering alone.~An effect appeared as if
there were an additional scattering mechanism activated at
temperatures above $200$~K which renormalized the acoustic
deformation-potential coupling constant so that it increased by a
factor of approximately two in the narrow temperature range
$200\!-\!355$~K.~In Ref.~\cite{Na00}, an extraordinary broadening
of the Ps peaks in MgF$_{2}$ was interpreted by involving
short-wavelength acoustic phonon scattering in terms of the
Umklapp mechanism. However, later on this mechanism was shown not
to be the case up to $T\sim\!10^{4}$~K~\cite{Bo01}, being ruled
out by the Boltzmann energy distribution of delocalized Ps atoms.

In the present Letter, we report the results of the analysis of
the temperature broadening of the momentum distribution of
delocalized Ps in MgF$_{2}$ in terms of the optical
deformation-potential scattering model (long-wavelength optical
phonons).~Such a nonpolar optical scattering mechanism is known to
be the case in crystals with two and more atoms per elementary
cell at elevated temperatures when a corresponding coupling
constant is non-zero because of the selection rules dictated by
local symmetry restrictions~\cite{Ridley,Bir}.~We obtain the
numerical value of the optical deformation-potential coupling
constant in MgF$_{2}$ for the first time and discuss possible
applications of delocalized Ps states for the study of symmetry
dependent structural properties of crystalline dielectrics.

In terms of the Green functions formalism~\cite{Mahan}, the linear
projection experimentally measured of the momentum distribution
(1D-ACAR spectrum) of the thermalized Ps atom interacting with
phonons at finite temperatures is given by~\cite{Na00}
\begin{eqnarray}
N_{1D}(p_{x})&\sim&\int_{-\infty}^{\infty}\!\!\!dp_{z}\int_{-\infty}^{\infty}\!\!\!dp_{y}
\int_{0}^{\infty}\!d\omega\;e^{-\textstyle\omega/k_{B}T}\nonumber\\
&\times&{\Gamma_{\mathbf{k}}(\omega)\over{(\omega-\textbf{p}^{2}/2M^{\ast})^{2}+
\Gamma^{2}_{\mathbf{k}}(\omega)}}\,, \label{Np}
\end{eqnarray}
\noindent where the exponential factor stands for the Boltzmann
statistics because there is at most only one Ps atom at a time
under usual experimental conditions.~The non-exponential factor
represents the so-called spectral density function in its explicit
form with $\Gamma_{\mathbf{k}}(\omega)$ being the imaginary
self-energy of Ps with the quasimomentum
$\textbf{k}\!=\!\textbf{p}/\hbar$. In the weak phonon coupling
regime (delocalized Ps --- see Ref.~\cite{Bo98}),
$\Gamma_{\mathbf{k}}(\omega)$ is usually written to the leading
(second) order approximation in the Ps interaction with a phonon
field,
\begin{eqnarray}
\Gamma_{\mathbf{k}}(\omega)&=&\pi\sum_{\mathbf{q}}|V_{\mathbf{q}}|^{2}
\{n(\omega_{\mathbf{q}})\,\delta(\omega-E_{\mathbf{k}+\mathbf{q}}+\hbar\omega_{\mathbf{q}})\nonumber\\
&+&[n(\omega_{\mathbf{q}})+1]\,\delta(\omega-E_{\mathbf{k}+\mathbf{q}}-\hbar\omega_{\mathbf{q}})\}\,,
\label{Gk}
\end{eqnarray}
where $V_{\mathbf{q}}$ is the Ps--phonon interaction matrix
element, $E_{\mathbf{k}}\!=\!\hbar^{2}\mathbf{k}^{2}/2M^{\ast}$ is
the energy of Ps with the band mass $M^{\ast}$,
$n(\omega_{\mathbf{q}})\!=\![\exp(\hbar\omega_{\mathbf{q}}/k_{B}T)-1]^{-1}$
is the equilibrium phonon distribution function, and
$\omega_\mathbf{q}$ is the frequency of the phonon with the wave
vector $\mathbf{q}$.~With allowance made for acoustic
deformation-potential scattering alone, Eq.~(\ref{Gk}) transforms
to~\cite{Na00,Bo01}
\begin{equation}
\Gamma_{\mathbf{k}}(\omega)=\Gamma^{(a)}_{\mathbf{k}}(\omega)=
{E_{d}^{\,2}M^{\ast\,3/2}k_{B}T\over{\sqrt{2}\,\pi\,\hbar^{3}u^{2}\rho}}
\,\sqrt{\omega}\,\,,\label{Ga}
\end{equation}
\noindent where $E_{d}$ is the coupling constant of Ps to the
defor\-mation-potential produced by long-wavelength longitudinal
acoustic vibrational modes, $u$ and $\rho$ are the sound velocity
and density of a crystal, respectively.

Magnesium Fluoride is an optically transparent insulator which
crystallizes with the rutile structure of the point group
D$_{4h}$~\cite{Hand}.~A group character analysis shows that this
material should exhibit nine optical infrared vibrational modes in
total~\cite{Barker} --- three nondegenerate longitudinal and three
doubly degenerate transverse modes of symmetry $E_{u}$ plus one
longitudinal and two nondegenerate transverse modes of symmetry
$A_{2u}$.~In our analysis, we apply an isotropic approximation in
considering nonpolar optical phonon scattering of Ps.~Within this
approximation the nonpolar optical contribution to the imaginary
self-energy (\ref{Gk}) is of the form~\cite{Bo01}
\begin{eqnarray}
\Gamma^{(o)}_{\mathbf{k}}(\omega)\!\!&=&\!\!{D_{o}^{2}M^{\ast\,3/2}\sqrt{\omega}
\over{2\,\sqrt{2}\,\pi\,\hbar^{2}\rho\,\omega_{o}}}
\left\{n(\omega_{o})\sqrt{1+{\hbar\omega_{o}\over{\omega}}}\right.\nonumber\\
\!\!&+&\!\!\left.[n(\omega_{o})+1]\;\theta\!\left(\!1-{\hbar\omega_{o}\over{\omega}}\right)\!
\sqrt{1-{\hbar\omega_{o}\over{\omega}}}\,\right\}, \label{Gof}
\end{eqnarray}
\noindent where $D_{o}$ is the coupling constant of Ps to the
defor\-mation-potential produced by long-wavelength optical
vibrations of averaged frequency $\omega_{o}$ with phonon
distribution function
$n(\omega_{o})\!=\![\exp(\hbar\omega_{o}/k_{B}T)-1]^{-1}$,
$\theta(x)$ is the unit-step function.~The total imaginary
self-energy accounting for two types of Ps-phonon scattering
(acoustic and nonpolar optical) is then written as
\begin{equation}
\Gamma_{\mathbf{k}}(\omega)=\Gamma^{(a)}_{\mathbf{k}}(\omega)+\Gamma^{(o)}_{\mathbf{k}}(\omega)
={\widetilde{E}_{d}^{\,2}(\omega)\,M^{\ast\,3/2}\,k_{B}T
\over{\sqrt{2}\,\pi\,\hbar^{3}u^{2}\rho}}\,\sqrt{\omega}
\label{Gt}
\end{equation}
with
\begin{eqnarray}
\widetilde{E}_{d}(\omega)\!\!&=&\!\!\!\left\{E_{d}^{\,2}+
{\hbar\,u^{2}D_{o}^{2}\over{2\,k_{B}T\,\omega_{o}}}\!\left[
n(\omega_{o})\sqrt{1+\!{\hbar\omega_{o}\over{\omega}}}\right.\right.\nonumber\\
\!\!&+&\!\!\!\left.\left.[n(\omega_{o})+1]\;\theta\!\left(\!1-{\hbar\omega_{o}\over{\omega}}\right)\!
\sqrt{1-{\hbar\omega_{o}\over{\omega}}}\,\right]\right\}^{\!\!1/2}
\label{Edw}
\end{eqnarray}
\noindent representing the 'effective' deformation-potential
coupling constant with nonpolar optical scattering taken into
account.~In view of an obvious fact that
$\omega\!\sim\!k_{B}T$~only contribute to the Ps momentum
distribution (\ref{Np}), one may approximate
$\widetilde{E}_{d}(\omega)\!\approx\!\widetilde{E}_{d}(k_{B}T)$
and then $\widetilde{E}_{d}$ is easily seen to increase in its
value from $E_{d}$ (for $T\!\ll\!\hbar\omega_{o}/k_{B}$) to
$\widetilde{E}_{d}\!=\!\sqrt{E_{d}^{\,2}+(uD_{o}/\omega_{o})^{2}}\,$
(for $T\!\gg\!\hbar\omega_{o}/k_{B}$).~This explains the increase
of the Ps--phonon coupling constant and corresponding
extraordinary broadening of the Ps momentum distribution in
MgF$_{2}$ at elevated temperatures reported earlier in
Ref.~\cite{Na00}.

Inserting Eqs.~(\ref{Gt}) and (\ref{Edw}) into Eq.~(\ref{Np}) and
performing integration over $p_{y}$ and $p_{z}$, one obtains in
dimensionless variables
\begin{eqnarray}
N_{1D}(p_{x})\!\!&\sim&\!\!\int_{0}^{\infty}\!\!\!d\xi\;\xi\;e^{\textstyle-\xi^{2}}\nonumber\\
\!\!&\times&\!\!\left\{\arctan\!\left[
{\xi^{2}-p_{x}^{2}/(2M^{\ast}k_{B}T)\over{\textstyle\Gamma^{(a)}(\xi)+
\Gamma^{(o)}(\xi)}}\right]+{\pi\over{2}}\right\} \label{Npx}
\end{eqnarray}
with $\,\xi\!=\!\sqrt{\hbar\omega/k_{B}T}\,$ and
$\,\Gamma^{(a),(o)}(\xi)\!=\!\Gamma^{(a),(o)}_{\bf
k}(\omega)/k_{B}T\,$. We use Eq.~(\ref{Npx}) to observe how
important nonpolar optical scattering is in MgF$_{2}$ at different
temperatures.~To this end we define a mean-square deviation
function
\begin{equation}
S=\sum_{i}\left\{\!{I_{1D}[p_{x}^{(i)}]\over{I_{1D}(0)}}-
{\overline{N}_{1D}[p_{x}^{(i)}]\over{\overline{N}_{1D}(0)}}\!\right\}^{\!\!2},
\label{chi2}
\end{equation}
where $\{I_{1D}[p_{x}^{(i)}]\}_{i=1}^{N}$ is a set of intensities
representing a~1D-ACAR spectrum measured at fixed temperature (a
complete description of experimental measurements is found
elsewhere~\cite{Na00}), $N$ is the number of data points,
$\overline{N}_{1D}$ is the momentum distribution (\ref{Npx})
convoluted with an experimental resolution function which was the
Gaussian function of $\mbox{FWHM}=0.297\times10^{-3}mc$ ($m$ is
the~free electron mass, $c$ is the speed of light).~We then plot
Eq.~(\ref{chi2}) with $E_{d}\!=\!7.6$~eV and
$M^{\ast}\!=1.1\times2m$~\cite{Na00} as a function of the optical
deformation-potential coupling constant $D_{o}$ at those
temperatures at which the ACAR measurements were done.~Other
material parameters were fixed as follows:
$\omega_{o}\!=\!7.7\!\times\!10^{13}$~rad/s,
$u\!=\!7\!\times\!10^{5}$~cm/s (averaged over directions and
phonon polarizations; from Refs.~\cite{Barker} and~\cite{Al69},
respectively), $\rho=3.13$~g/cm$^{3}$~\cite{Hand}.

\begin{figure}[b]
\epsfxsize=8.66cm\centering{\epsfbox{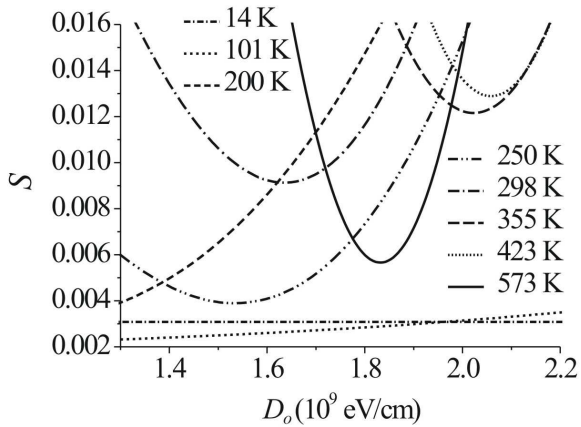}}\vspace{-0.5cm}
\caption{The function $S(D_{o})$ at different temperatures.}
\label{fig1}
\end{figure}

\begin{figure}[t]
\epsfxsize=5.65cm\centering{\epsfbox{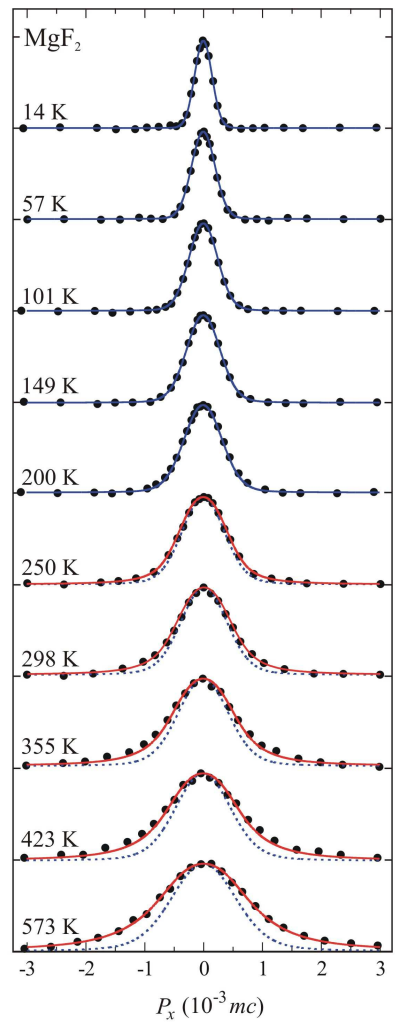}} \caption{(Color
online) The central peaks of the 1D-ACAR spectra of Ps in
MgF$_{2}$ at different temperatures as calculated from
Eq.(\ref{Npx}) convoluted with an experimental resolution function
(see text for explanations).} \label{fig2}
\end{figure}

The results are presented in Fig.~\ref{fig1}.~At $T\!\le\!200$~K,
the function~$S(D_{o})$ is either a constant (at very low
$T\!\sim\!10$~K) or tends to its minimal value at
$D_{o}\!\sim\!0$,~thus indicating the absence of nonpolar optical
scattering at these temperatures.~This is clear as optical
vibrational degrees of freedom are frozen at temperatures below
$\hbar\omega_{o}/2k_{B}$ (the optical zero-point vibration energy
devided by $k_{B}$) which is estimated to be $\sim\!300$~K
for~MgF$_{2}$.~For all $T\!\ge\!250$~K, clear minima are seen in
$S(D_{o})$ at approximately the same value of the optical
deformation-potential coupling constant
$\sim\!1.8\!\times\!10^{9}$~eV/cm. This indicates that nonpolar
optical scattering comes into play above 250~K and starts
broadening the Ps momentum distribution which was initially
broadened by acoustic deformation-potential scattering. From this
analysis we obtain $D_{o}=(1.8\pm0.3)\times10^{9}$~eV/cm for the
delocalized Ps atom in MgF$_{2}$.~Figure~\ref{fig2} shows the
theoretical curves $\overline{N}_{1D}(p_{x})/\overline{N}_{1D}(0)$
(solid lines) plotted for this value of $D_{o}$ along with the
normalized central peaks of the experimental 1D-ACAR spectra at
different temperatures. The dashed lines above 200~K represent the
theoretical momentum distribution with only acoustic
deformation-potential scattering taken into account. The
experimental spectra are clearly seen to be very nicely reproduced
theoretically at all the temperatures in terms of the Ps nonpolar
optical scattering model.

A similar effect of an optical deformation-potential was not
observed in $\alpha$-SiO$_{2}$ (crystalline quartz) where the
temperature broadening of the central and satellite 1D-ACAR peaks
of delocalized Ps was explained in terms of acoustic
deformation-potential scattering alone throughout the entire
temperature range $88-701$~K~\cite{Na00}.~In long-wavelength
optical vibrations, one set of atoms moves as a body against the
second set of atoms of the same elementary cell.~This changes the
particle's energy by an amount
$\sim\!\mathbf{D}_{o}\!\cdot\!\mathbf{u}$ with $\mathbf{u}$ being
the relative displacement of the two atomic sets and
$\mathbf{D}_{o}$ representing the vectorial optical
deformation-potential constant ($|\mathbf{D}_{o}|\!=\!D_{o}$)
which is nothing but the matrix element of a perturbation operator
(i.e.~the variation of the rigid-lattice potential due to optical
vibrations) taken over Bloch wave-functions of the particle in the
neighborhood of its energy band minimum~\cite{Ridley,Bir}.~Since
we deal with the Ps atom having a parabolic band with its minimum
in the center of the Brillouin zone ($\Gamma$-valey), for this
additional energy to be non-zero, $\mathbf{D}_{o}$ and
$\mathbf{u}$ must transform according to equal-dimension
irreducible representations of the point symmetry group of the
$\Gamma$-point of the Brillouin zone.~In terms of such a symmetry
analysis~\cite{Bir}, optical deformation-potential scattering is
allowed in the $\Gamma$-valley of the MgF$_{2}$ lattice (point
group $D_{4h}$) and forbidden in the $\Gamma$-valley of the
$\alpha$-SiO$_{2}$ lattice (point group $D_{3}$ with the
$C_{3}$-axis being a screw axis).~However, crystalline quartz is
known to undergo the second-order structural phase transition from
$\alpha$- to $\beta$-phase above 846~K~\cite{Hand} where its local
symmetry changes from $D_{3}$ to $D_{6}$.~The latter one is
isomorphic (thus having the same set of group representations and
the same selection rules for deformation-potential scattering in
the $\Gamma$-valey) to the $C_{6v}$ local symmetry of
wurtzite-type crystals which was analyzed in detail in
Ref.~\cite{Bir} and shown to allow optical deformation-potential
scattering in the $\Gamma$-valey.~Thus, Ps nonpolar optical phonon
scattering must manifest itself in $\beta$-SiO$_{2}$ by
drastically broadening the Ps 1D-ACAR peaks above
846~K~\cite{Bo04}.

In summary, temperature broadening of the momentum distribution of
delocalized Ps in Magnesium Fluoride has been analyzed in terms of
optical deformation-potential scattering model.~The Ps optical
deformation-potential coupling constant $D_{o}$ in MgF$_{2}$ has
been determined for the first time to be
$(1.8\pm0.3)\times10^{9}$~eV/cm.~In view of the fact that the
lowest positron band in alkali halides is not considerably
different from the conduction band of the electron~\cite{Kunz},
the positron optical deformation-potential coupling constant may
be thought of being approximately equal to that of the
electron.~Then, one obtains an estimate of $9\times10^{8}$~eV/cm
for the electron optical deformation-potential coupling constant
in MgF$_{2}$.~This agrees with typical values usually given in the
literature for other materials (see, e.g.~\cite{Neuberger}).~We
have also shown that, as short-range deformation-potentials are
essentially symmetry dependent, Ps may be able to sense symmetry
dependent structural properties of dielectric crystals.~In
particular, high-temperature second-order phase transitions, where
the local symmetry of a crystal lattice changes but the crystal
structure remains, may be identified via jumps in the temperature
dependences of HWHMs of Ps peaks measured in ACAR-experiments in
those dielectric crystals where optical deformation-potential
scattering is allowed in one and forbidden in another crystalline
phase.

\begin{acknowledgments}
One of the authors (I.B.) would like to acknowledge a~financial
support from the University of Savoie (France) within a~Visiting
Professorship Program.~I.B. would also like to thank
I.D.~Feranchuk, L.V.~Keldysh, L.I.~Komarov, and V.N.~Kushnir for
useful discussions.
\end{acknowledgments}

\end{document}